\documentclass[final,1p,times]{elsarticle}

\usepackage{amssymb}
\usepackage{amsmath}

\usepackage[utf8]{inputenc}
\usepackage[T1]{fontenc}
\usepackage[colorlinks,allcolors=blue]{hyperref}
\usepackage{float}
\usepackage{multirow}
\usepackage{tabularx}
\usepackage{array}

\graphicspath{ {./images/} }

\journal{Software Impacts}

\begin{document}

\begin{frontmatter}



\title{Retuve: Automated Multi-Modality Analysis of Hip Dysplasia with Open Source AI}


\author[alberta]{Adam McArthur}
\author[alberta]{Stephanie Wichuk}
\author[alberta]{Stephen Burnside}
\author[nhs lothian]{Andrew Kirby}
\author[isc]{Alexander Scammon}
\author[isc]{Damian Sol}
\author[alberta]{Abhilash Hareendranathan}
\author[alberta]{Jacob L. Jaremko}

\affiliation[alberta]{Department of Radiology and Diagnostic Imaging, University of Alberta, Edmonton, AB T6G 2R3, Canada}
\affiliation[nhs lothian]{Department of Radiology, NHS Lothian, Edinburgh, EH16 4SA, United Kingdom}
\affiliation[isc]{Insight Softmax Consulting, 4 Embarcadero Center, Suite 1400 San Francisco, CA 94111}

\begin{abstract}
Developmental dysplasia of the hip (\textbf{DDH}) poses significant diagnostic challenges, hindering timely intervention. Current screening methodologies lack standardization, and AI-driven studies suffer from reproducibility issues due to limited data and code availability. To address these limitations, we introduce Retuve, an open-source framework for multi-modality \textbf{DDH} analysis, encompassing both ultrasound (\textbf{US}) and X-ray imaging. Retuve provides a complete and reproducible workflow, offering open datasets comprising expert-annotated \textbf{US} and X-ray images, pre-trained models with training code and weights, and a user-friendly Python Application Programming Interface (\textbf{API}). The framework integrates segmentation and landmark detection models, enabling automated measurement of key diagnostic parameters such as the alpha angle and acetabular index. By adhering to open-source principles, Retuve promotes transparency, collaboration, and accessibility in \textbf{DDH} research. This framework can democratize \textbf{DDH} screening, facilitate early diagnosis, and improve patient outcomes by enabling widespread screening and early intervention. The GitHub repository/code can be found here: \url{https://github.com/radoss-org/retuve}
\end{abstract}

\begin{graphicalabstract}
\includegraphics[width=\linewidth]{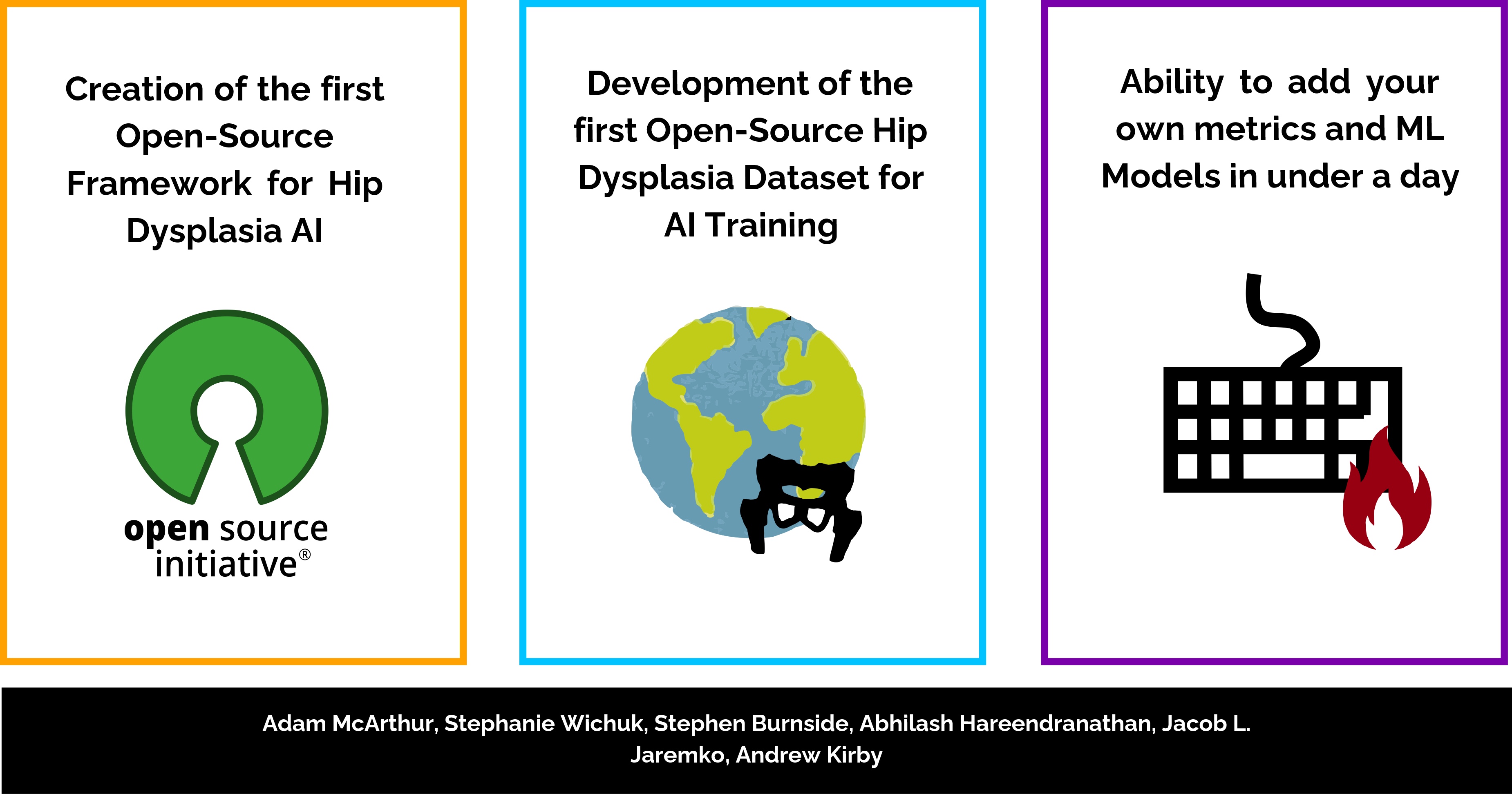}
\end{graphicalabstract}

\begin{highlights}
\item Creation of an open-source framework facilitating ongoing research in \textbf{DDH} imaging, promoting collaborative advancement in the field.

\item Development and release of a pioneering \textbf{DDH} open-source dataset, complete with expert annotations for both ultrasound and X-ray imaging modalities.

\item Implementation of a modular AI system that seamlessly integrates segmentation and landmark models into Retuve, with three published plugins demonstrating its versatility.
\end{highlights}


\end{frontmatter}



\section*{Nomenclature}

\begin{table}[H]
    \centering
    \begin{tabular}{|l l|}
    \hline
    \textbf{Acronym} & \textbf{Definition} \\
    \hline
    \textbf{AI}   & Artificial Intelligence \\
    \textbf{DDH}  & Developmental Dysplasia of the Hip \\
    \textbf{FP}   & False Positive \\
    \textbf{GHDR} & Global Hip Dysplasia Registry \\
    \textbf{ICML} & International Conference on Machine Learning \\
    \textbf{MICCAI} & Medical Image Computing and Computer Assisted Intervention \\
    \textbf{OSI}  & Open Source Initiative \\
    \textbf{PACS} & Picture Archiving and Communication System \\
    \textbf{POCUS} & Point-of-Care Ultrasound \\
    \textbf{RCT}  & Randomized Controlled Trial \\
    \textbf{US}   & Ultrasound \\
    \textbf{VicHip} & Victorian Hip Dysplasia Registry \\
    \textbf{MIMIC-III} & The Medical Information Mart for Intensive Care III \\
    \textbf{LLM} & Large Language Model \\
    \textbf{API} & Application Programming Interface \\
    \textbf{MRI} & Magnetic Resonance Imaging \\
    \textbf{CT} & Computed Tomography \\
    \textbf{GUI} & Graphical User Interface \\
    \textbf{CLI} & Command Line Interface \\
    \textbf{CPU} & Central Processing Unit \\
    \textbf{RAM} & Random Access Memory \\
    \textbf{FPS} & Frames Per Second \\
    \textbf{ICC} & Intraclass Correlation Coefficient \\
    \hline
    \end{tabular}
    \caption{List of acronyms used in this paper.}
    \label{tab:nonclementure}
\end{table}

\section{Introduction}
Developmental dysplasia of the hip (\textbf{DDH}) research is hampered by the absence of standardized, objective screening methodologies and limitations conducting robust, reproducible AI-driven studies.

The definition threshold for \textbf{DDH} is unclear. There are high costs to universal ultrasonographic screening programs, varying reported incidence rates from 1.5 to 20 per 1000 births \cite{Starr2014}, and inconsistent risk factor assessments, such as breech position carrying a 2-24 times increased risk of \textbf{DDH} \cite{breech-bache2002risk, breech-de2012risk, breech-kotlarsky2015developmental, breech-schwend2014evaluation, breech-swarup2018developmental}.

Despite high sensitivity rates of 88-95\% for well-performed ultrasound screening \cite{DEGNAN2021213}, the condition faces challenges typical of rare diseases. Recent studies report that False Positive (\textbf{FP}) cases are costly for screening and occur at high rates \cite{ukppv2017}. Most immature hips self-resolve - 89--98\% of Graf IIa, \~90\% of Graf IIb, and even some Graf III cases normalize without intervention \citep{Theunissen2025,Sakkers2018NaturalHistory}, while global screening protocols lack consensus on who to screen, when and how \cite{kilsdonk2021ultrasound}.

Long-term randomized controlled trials (\textbf{RCTs}) extending into later life are needed to validate screening and treatment protocols \cite{kilsdonk2021ultrasound}. Multiple large-scale efforts are gathering data for these \textbf{RCT}s, such as the Global Hip Dysplasia Registry (\textbf{GHDR}) \cite{GHDR2025} and the Victorian Hip Dysplasia Registry (\textbf{VicHip}) \cite{VicHip2025}.

Open data is vital for advancing reproducibility, transparency, and innovation in open-source AI research, particularly in medicine, where reliability and generalizability are critical. The reproducibility crisis in AI, driven by limited access to datasets, code, and methodologies, highlights the need for open data. Studies show that less than a third of AI research is reproducible, with inaccessible healthcare datasets being a major barrier \cite{6b_ai_reproducibility,9b_ai_research_transparency,14b_covid_open_data}. Only 26\% of AI studies in a systematic review were computationally reproducible, rising to 86\% when both code and data were shared \cite{9b_ai_research_transparency}. In medical AI research, up to 97\% of studies lacked sufficient transparency for real-world application \cite{16b_ai_transparency_healthcare}.

Open data initiatives like \textbf{MIMIC-III} and OpenML have improved reproducibility by enabling validation across diverse populations and settings. \textbf{MIMIC-III} provides de-identified health records for over 40,000 patients, fostering collaboration and reproducibility in clinical research \cite{5b_mimic_analysis_platform, 22b_mimiciii_database}. OpenML facilitates sharing of machine learning datasets and experiments, promoting transparent benchmarking and algorithm development \cite{17b_openml_repository, 19b_openml_task_example}. Leading conferences like \textbf{MICCAI} and \textbf{ICML} emphasize open data's value. \textbf{MICCAI} 2024 launched an Open Data initiative focusing on underrepresented populations to address global healthcare challenges \cite{39b_miccai_open_data_2024, 50b_miccai_news_open_data}. \textbf{ICML} promotes "data-centric AI," highlighting the importance of high-quality datasets for impactful research \cite{54b_icml_dmlr_2024}.

Openly sharing data improves the quality, reliability, and usefulness of AI research in medicine. This approach mitigates biases and accelerates innovation by enabling researchers worldwide to validate and build upon existing work \cite{1b_open_access_healthcare,11b_clinical_data_research,24b_open_data_healthcare}.

We present Retuve, an open-source framework for \textbf{DDH} analysis of ultrasound and X-ray imaging that adheres to the \textbf{OSI} definition of open-source AI \cite{OpenSourceAI2024}. Retuve provides complete reproducibility, offering open datasets and models (training code and weights), and a Python Application Programming Interface (\textbf{API}). This commitment to transparency and reproducibility empowers researchers and clinicians to collaboratively advance automated \textbf{DDH} detection. Retuve's white-box algorithm allows detailed analysis and improvement of AI models for ultrasound and X-ray screening. Retuve will expand its open-source data and AI literature coverage over the next 2-3 years, fostering collaborative efforts to advance \textbf{DDH} understanding and treatment. Retuve aligns with similar efforts in the \textbf{LLM} community to provide fully-open source AI \cite{olmo2}. Retuve aims to support widespread screening and early intervention for \textbf{DDH}, potentially reducing reliance on highly skilled professionals, facilitating earlier diagnosis, and preventing early-onset osteoarthritis linked to \textbf{DDH}.

\begin{figure}[H]
    \centering
    \includegraphics[width=\textwidth]{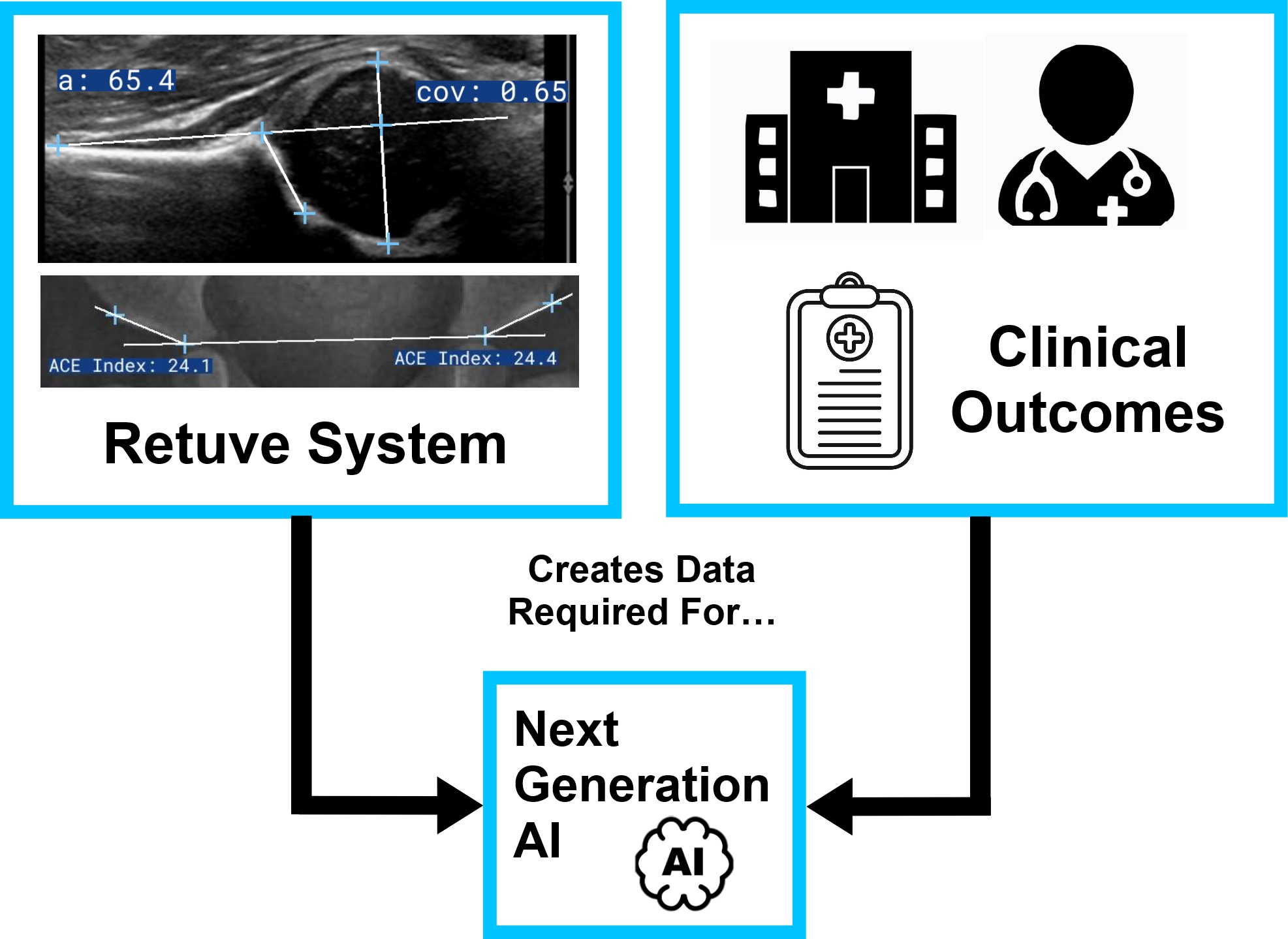}
    \caption{Retuve can analyse ultrasounds and X-rays to generate conventional and new indices quantifying hip anatomy, such as acetabular alpha angle and coverage (ultrasound), and acetabular index (X-ray). These indices can be used to train the next generation of AI Models to diagnose \textbf{DDH}.}
    \label{fig:retuve}
\end{figure}

We hope to combine Retuve with Clinical Outcomes being acquired at \textbf{GDHR} and \textbf{VicHip} to create comprehensive AI models that can detect and risk-stratify \textbf{DDH} in diverse settings, as shown in \textbf{Figure \ref{fig:retuve}}. This includes automatic classification of Graf class (normal to severely dysplastic) and scan quality.

\section{Research Question}

\textbf{Research Question:} Can we develop and validate a comprehensive, modular open-source framework that addresses the reproducibility crisis in \textbf{DDH} AI research by providing complete multi-modality analysis capabilities (ultrasound and X-ray), standardized workflows, open datasets and models, while maintaining sufficient performance for clinical applications and enabling community-driven collaborative advancement through extensible architecture?

\section{Code Metadata}

\begin{table}[H]
    \centering
    \begin{tabular}{|p{0.4\linewidth}|p{0.5\linewidth}|}
    \hline
    \textbf{Item} & \textbf{Description} \\
    \hline
    Current code version & v0.1.0 \\
    \hline
    Permanent link to code/repository & \url{https://github.com/radoss-org/retuve} \\
    \hline
    Latest commit at time of Publication & 89a53e923a7b907d4e772ed029464866002ee2a6 \\
    \hline
    Legal Code License & Apache V2.0 Licence \\
    \hline
    Code versioning system used & git \\
    \hline
    Project Requirements & Please see the \href{https://github.com/radoss-org/retuve/blob/main/pyproject.toml}{project.toml file} \\
    \hline
    Support Protocol & Email adam@mcaq.me or create a GitHub Issue \\
    \hline
    \end{tabular}
    \caption{Software metadata}
    \label{tab:metadata}
\end{table}

\section{Related Work}

\subsection{Ultrasound}

\textbf{DDH} diagnosis increasingly leverages Artificial Intelligence (AI) techniques, yet the field suffers from lack of consensus and standardization across methodologies. While some studies suggest that identifying alpha/beta angles and coverage on a single ultrasound frame is straightforward \cite{7950680, HAREENDRANATHAN201689, 9448429, clement2024, 9486886}, this perspective overlooks the complexities of acquiring standardized images. Accurate diagnosis often necessitates a standard-plane detector and analysis of full "2D sweep videos" to capture the dynamic nature of the hip joint \cite{Kinugasa2023, diagnostics12061423, jaremko2023aimedo}. Research into 3D ultrasound \cite{hareendranathan2022normal, GHASSEMINIA2022101082, HAREENDRANATHAN201689} aims to extract measurements beyond the standard plane for comprehensive assessment of hip morphology.

Strategies for analyzing ultrasound images vary considerably. These approaches can be categorized into contour methods and landmark methods. Contour methods use segmentations to draw a "contour" of the ilium, acetabulum and femoral head to find landmarks \cite{jaremko2023aimedo, 7950680, HAREENDRANATHAN201689}. Landmark methods skip this step by finding landmarks directly, but often require senior-expert labels and are less robust to poor quality scans \cite{jan2023assessing, 9449886}.

Recognizing the importance of image quality in accurate diagnosis, researchers have successfully applied AI to predict scan quality and develop scoring systems for ultrasound image quality \cite{Hareendranathan2021, Hareendranathan2022, 9449886, app12084072}. This is crucial for ensuring reliable and consistent AI-driven \textbf{DDH} assessments \cite{9448429, clement2024, diagnostics12061423, 9486886}.

\subsection{X-ray}

AI-based methods for X-ray analysis of \textbf{DDH} follow distinct approaches. One approach involves directly classifying a hip as Normal or Dysplastic. One study utilized YOLOv5 to directly predict dysplasia presence with an associated confidence score \cite{den2023diagnostic}. This end-to-end approach offers a potentially rapid screening tool.

Alternatively, Landmark methods are frequently employed. These methods typically identify key anatomical landmarks on the X-ray image to calculate various measurements, including the Acetabular Index. This information, combined with additional metadata, is then used to classify the hip as Normal or Dysplastic \cite{xraylandmarksflow2023}. Landmark detection accuracy is paramount for reliable classification.

Another approach involves segmenting the pelvis and femur, followed by applying a specialized algorithm to identify correct landmarks for calculating the acetabular angle \cite{jan2023assessing}. This segmentation-based approach can improve landmark detection robustness, particularly in cases with subtle anatomical variations.

\subsection{The Critical Gap: Absence of Open Source Standardization in DDH}

Despite growing research in AI-assisted \textbf{DDH} diagnosis, a significant gap remains in reproducibility and accessibility. Currently, no research paper on \textbf{DDH} AI provides fully reproducible code and data for the AI models used. While some papers include code snippets \cite{AutoDDH2025} or utilize open-source data with limited labels \cite{jan2023assessing, 9449886}, a concerted effort to standardize methodologies and provide comprehensive open-source resources is lacking. This absence hinders independent validation, comparison of different approaches, and further field development.

Retuve distinguishes itself from existing methods through its comprehensive approach to addressing challenges in \textbf{DDH} research, focusing on reproducibility, standardization, and accessibility. While previous studies have explored AI-driven analysis of ultrasound and X-ray images for \textbf{DDH} diagnosis, they often fall short in providing necessary resources for independent validation and further development.

\section{Retuve Overview}

Retuve is a comprehensive open-source framework for automated multi-modality \textbf{DDH} analysis, supporting ultrasound and X-ray imaging. It provides complete reproducible workflows with open datasets, models (training code and weights), and a Python API. The modular architecture enables seamless AI model integration for automated measurement of diagnostic parameters like alpha angle and acetabular index. Retuve promotes transparency and accessibility in \textbf{DDH} research while being adaptable for other medical imaging tasks including \textbf{MRI}/CT.

Retuve comprises two main components (\textbf{Figure \ref{fig:architecture}}):

\begin{figure}[H]
    \centering
    \includegraphics[width=\textwidth]{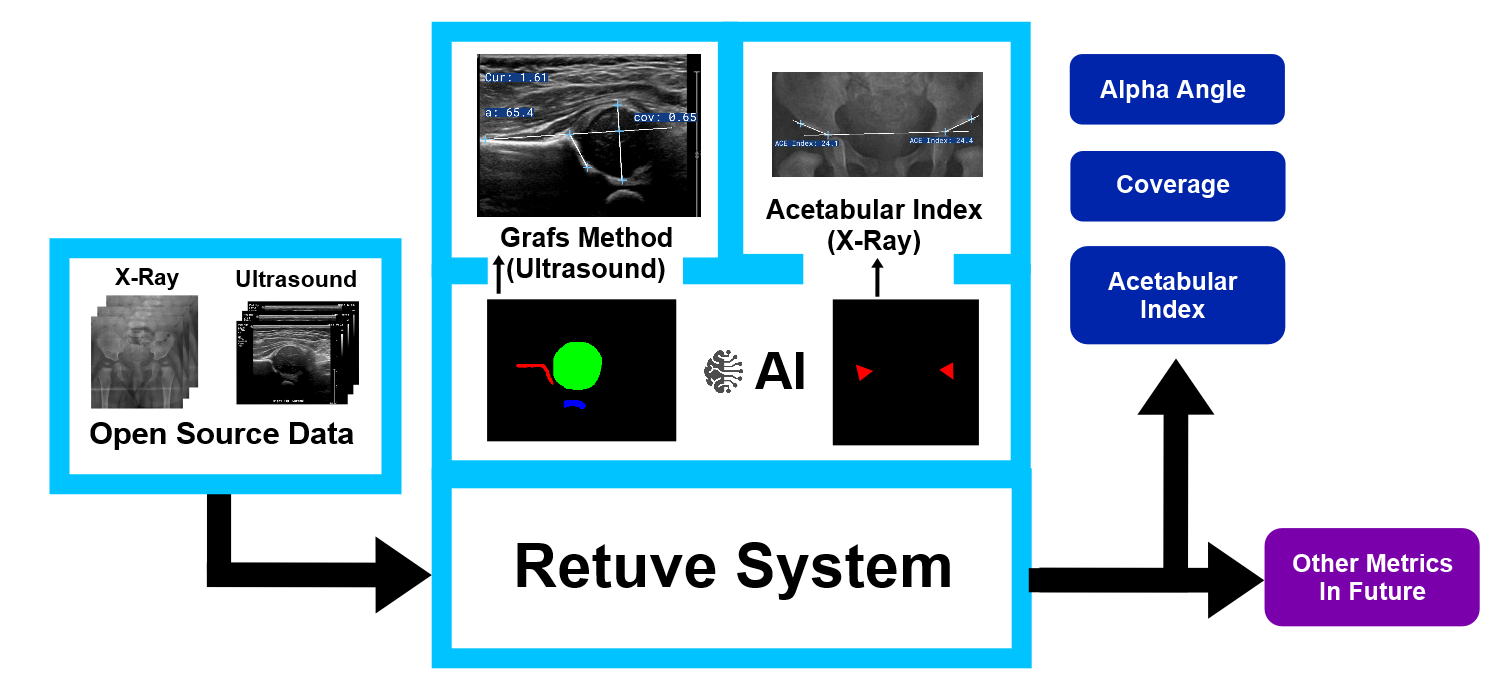}
    \caption{Retuve Architecture Diagram, with Inputs, AI Outputs, and final outputs for the user.}
    \label{fig:architecture}
\end{figure}

\begin{itemize}
    \item \textbf{Segmentation AI}: Neural network models that take the raw image and output the required geometry for analysis.
    \item \textbf{Rule-Based Algorithm}: Algorithms that take the geometry from the AI and output the final measurements.
\end{itemize}

Together, these components allow Retuve to analyze any of the above formats and output the required measurements.
The Segmentation AI System is shared between all modalities, and the Rule-Based Algorithm is specific to each modality.

\vspace{0.5cm}

\textbf{Retuve is being released before multi-center, clinical validation results have been published. Therefore all metrics have been marked as experimental when running Retuve. As metric validations are published, metrics will move through an alpha/beta/mature release system. The current release state and updates to the release plan can be found \href{https://github.com/radoss-org/retuve/milestones}{here}.}

\begin{figure}[h!]
    \centering
    \includegraphics[width=\textwidth]{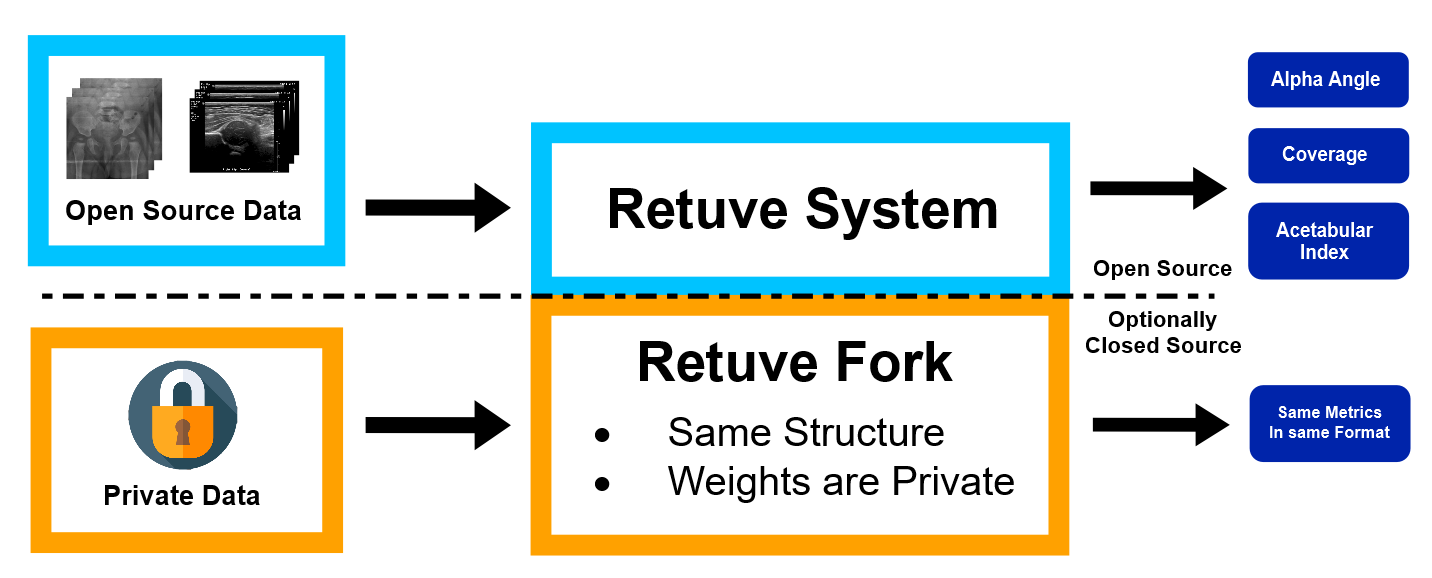}
    \caption{Retuve can easily be adjusted with a custom AI trained on any data, including forking to closed versions by those holding private data.}
    \label{fig:custom_ai}
\end{figure}

\subsection{AI Plugin System}

Following OSI open-source AI definitions \cite{OpenSourceAI2024}, Retuve supports segmentation models (instance and semantic) and landmark models. New models integrate with minimal effort (\textbf{Figure \ref{fig:custom_ai}}), enabling rapid upgrades and comparisons.

Retuve includes three AI architectures for ultrasound and X-ray analysis \cite{radiopaedia_ddh_cases, Chen_TinyUNet_MICCAI2024,ultralytics2025, isensee2018nnunetselfadaptingframeworkunetbased}:

\begin{itemize}
    \item \href{https://github.com/radoss-org/retuve-tinyunet-plugin}{retuve-tinyunet-plugin}
    \item \href{https://github.com/radoss-org/retuve-nnunet-plugin}{retuve-nnunet-plugin}
    \item \href{https://github.com/radoss-org/retuve-yolo-plugin}{retuve-yolo-plugin}
\end{itemize}

We recommend starting with the YOLO plugin, as it is the most well developed and tested.

\subsection{Rule-Based Algorithm}

Retuve employs two algorithms for different imaging modalities.

\subsubsection{Ultrasound Algorithm}

\begin{figure}[H]
    \centering
    \includegraphics[width=\textwidth]{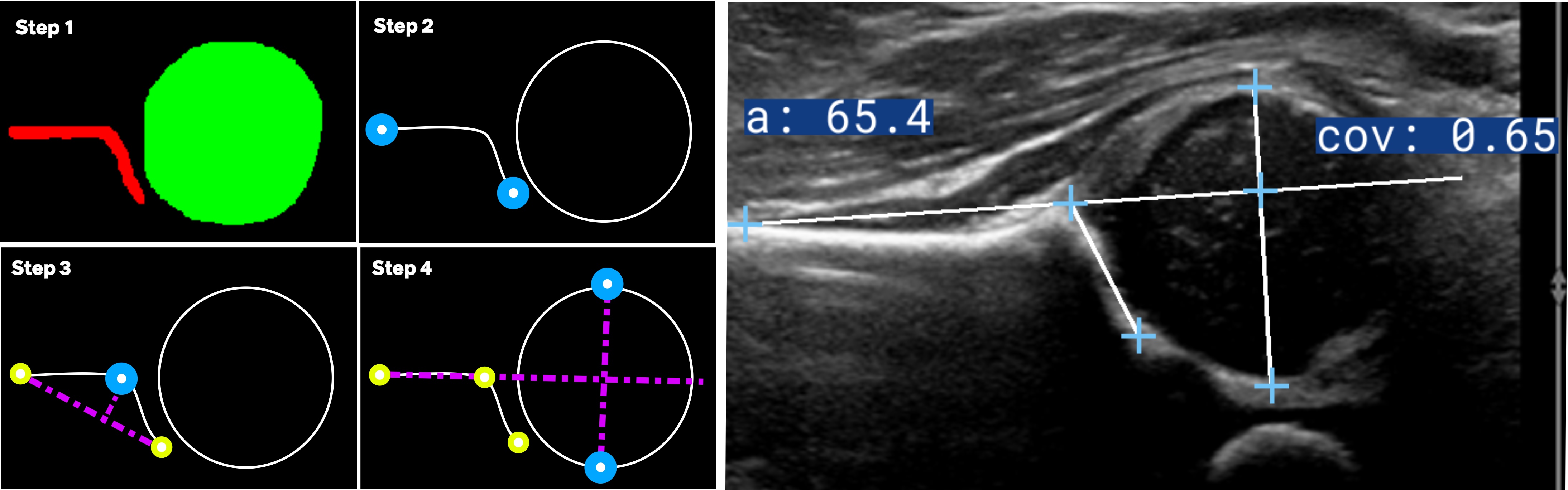}
    \caption{The ultrasound algorithm illustrated in 4 steps.}
    \label{fig:usalgo}
\end{figure}

Based on the Graf Method \cite{Graf1980}, this algorithm processes Ilium/Acetabulum and Femoral Head segmentations to generate five landmarks for Alpha Angle and Coverage calculations (\textbf{Figure \ref{fig:usalgo}}), building on established contour methods \cite{7950680, HAREENDRANATHAN201689}.

\subsubsection{X-Ray Algorithm}

\begin{figure}[H]
    \centering
    \includegraphics[width=\textwidth]{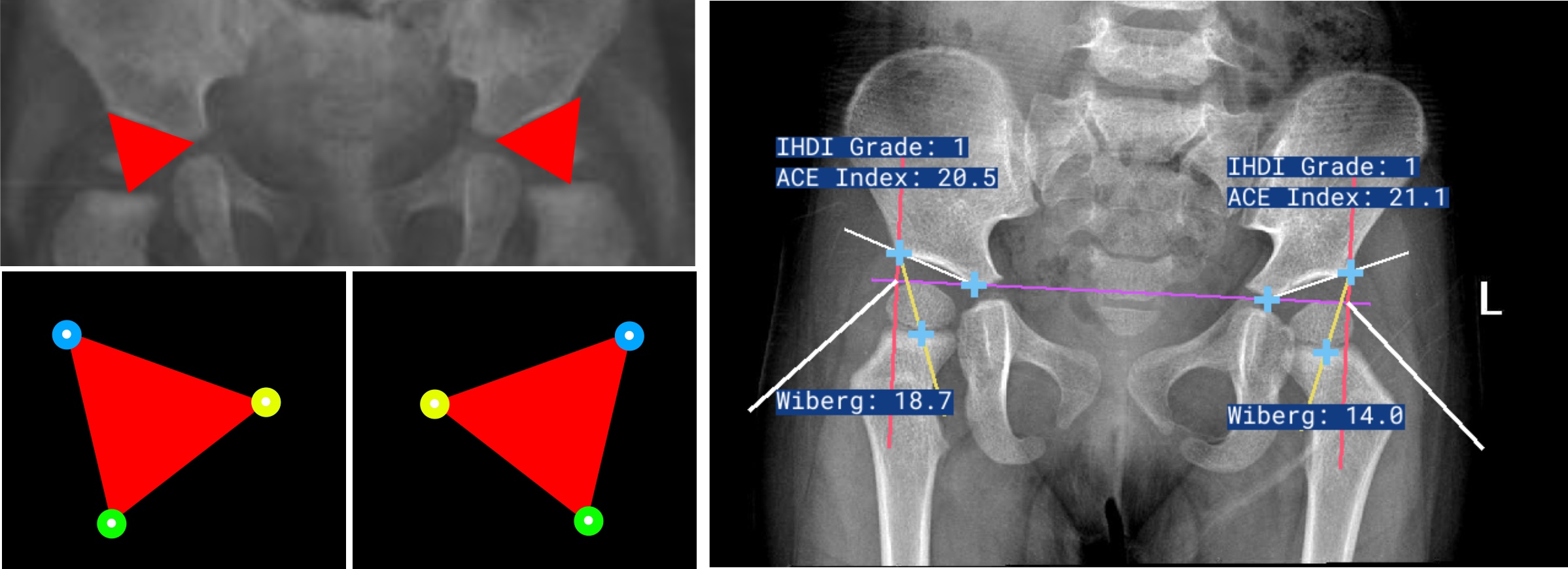}
    \caption{The algorithm for X-ray. Triangle segmentations are used to find the inner (yellow), outer (blue), and lower (green, h point) landmarks on each side of the pelvis. This allows us to calculate the Acetabular Index, Wilberg Index and IHDI grade.}
    \label{fig:xray_algo}
\end{figure}

This algorithm measures Acetabular Index and Wilberg Index using a "triangle" landmark method with three points (including the h point), enabling segmentation models to function as landmark models (\textbf{Figure \ref{fig:xray_algo}}). IHDI grade classification is derived from these measurements. Additional radiographic indices can be added as modular extensions.

\subsection{Open Source Data}

Retuve includes a curated dataset of 172 2D ultrasounds from Radiopedia \cite{radiopaedia_ddh_cases} and Hong Kong Polytechnic University \cite{9449886}, and X-ray data from the MTDDH dataset \cite{c088644bd0b2406eb49830ad447c17fb}, both available in the \href{https://github.com/radoss-org/open-hip-dysplasia}{Open Hip Dysplasia Dataset} \cite{openhipdysplasia}. \textbf{Note:} This represents the only good-quality Creative Commons licensed \textbf{DDH} imaging data currently available on the internet, highlighting the critical need for open datasets in this field.

\section{Software Impact/Results}

\subsection{Ease of Use}

\begin{figure}[H]
    \centering
    \includegraphics[width=\textwidth]{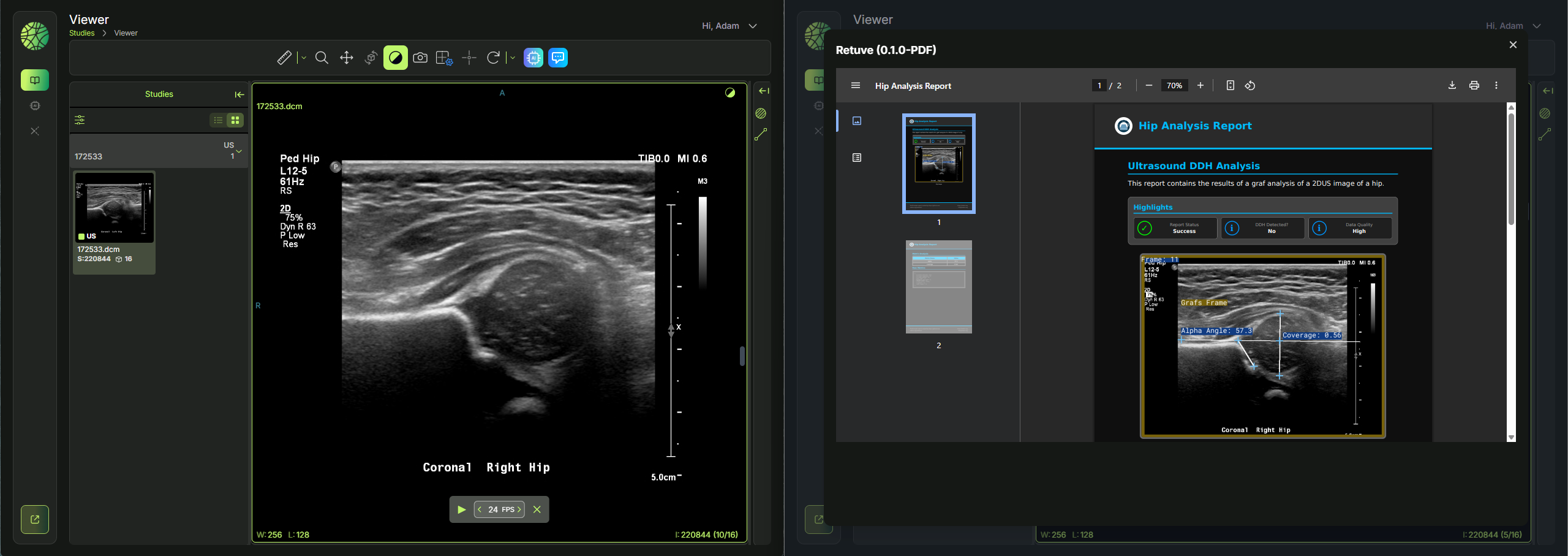}
    \caption{Retuve's integration with PACS-AI enables seamless deployment in clinical workflows without infrastructure overhaul. Shown left: the PACS-AI interface. Shown right: The DDH Report generated with the Retuve PACS-AI Plugin.}
    \label{fig:pacs_ai}
\end{figure}

Retuve enables complex \textbf{DDH} analysis with minimal code and integrates with PACS-AI \cite{PACSAI} for seamless clinical deployment (\textbf{Figure \ref{fig:pacs_ai}}). PACS-AI is an artificial intelligence deployment platform for medical imaging that enhances physician workflow with state-of-the-art AI models, seamlessly integrated with existing PACS infrastructure without requiring system overhaul.

\subsection{Validation Results}

Retuve's performance was evaluated using the \href{https://github.com/radoss-org/retuve-yolo-plugin}{retuve-yolo-plugin} for both ultrasound and X-ray modalities. All datasets were sourced from the \href{https://github.com/radoss-org/open-hip-dysplasia}{Open Hip Dysplasia Dataset} \cite{openhipdysplasia}. We report Intraclass Correlation Coefficient (\textbf{ICC}) values as our primary reliability measure, as \textbf{ICC} provides comprehensive assessment of both consistency and absolute agreement between AI-generated and expert measurements, making it the gold standard for evaluating measurement reliability in clinical applications.

\subsubsection{Dataset Information}

\textbf{Ultrasound}: 172 images with an 80/20 training split to ensure sufficient training data. A segmentation model was trained, followed by post-training validation to assess alpha angle and coverage accuracy.

\textbf{X-ray}: The dataset underwent systematic filtering from an initial collection of 2,156 images. A held-back test set from Keypoints Dataset 1 (122 images) was excluded to maintain experimental integrity. An additional 156 images were subsequently removed due to conversion failures during YOLO format preprocessing, resulting in a final dataset of 1,878 images suitable for analysis. Data was partitioned approximately equally with 937 images allocated for training and 941 for testing. Triangle segmentations were formed from anatomical points as described in the methodology.

\subsubsection{Ultrasound Performance}

\begin{table}[H]
    \centering
    \small
    \begin{tabularx}{\textwidth}{|>{\raggedright\arraybackslash}p{2.5cm}|>{\centering\arraybackslash}X|>{\centering\arraybackslash}p{2.5cm}|>{\centering\arraybackslash}X|>{\centering\arraybackslash}p{2.5cm}|}
    \hline
    \textbf{Metric} & \textbf{ICC Absolute} & \textbf{95\% CI} & \textbf{ICC Consistency} & \textbf{95\% CI} \\
    \hline
    Alpha Angle & 0.47 & -0.07 to 0.81 & 0.86 & 0.72 to 0.93 \\
    \hline
    Coverage & 0.88 & 0.80 to 0.96 & 0.92 & 0.84 to 0.96 \\
    \hline
    \end{tabularx}
    \caption{Ultrasound validation results for alpha angle and coverage measurements}
    \label{tab:ultrasound_validation}
\end{table}

\subsubsection{X-ray Performance}

\begin{table}[H]
    \centering
    \small
    \begin{tabularx}{\textwidth}{|>{\raggedright\arraybackslash}p{4cm}|>{\centering\arraybackslash}X|>{\centering\arraybackslash}X|}
    \hline
    \textbf{Metric} & \textbf{Left Side} & \textbf{Right Side} \\
    \hline
    ICC Acetabular Index & 0.860 (95\% CI 0.830-0.880) & 0.845 (95\% CI 0.810-0.870) \\
    \hline
    ICC Wilberg Index & 0.891 (95\% CI 0.860-0.910) & 0.902 (95\% CI 0.860-0.930) \\
    \hline
    \end{tabularx}
    \caption{X-ray validation results for acetabular index and Wilberg angle measurements}
    \label{tab:xray_validation}
\end{table}

\subsubsection{Classification Performance}

For X-ray \textbf{DDH} classification, Retuve demonstrated strong performance in distinguishing Grade 1 IHDI from Grades 2, 3, and 4:

\begin{table}[H]
    \centering
    \small
    \begin{tabularx}{\textwidth}{|>{\raggedright\arraybackslash}p{3cm}|>{\centering\arraybackslash}X|>{\centering\arraybackslash}X|>{\centering\arraybackslash}X|}
    \hline
    \textbf{Classification Task} & \textbf{F1 Score} & \textbf{Recall} & \textbf{Precision} \\
    \hline
    Grade 1 vs. Grades 2-4 IHDI & 0.940 & 0.914 & 0.967 \\
    \hline
    Per-Class (All Grades) & 0.593 & 0.570 & 0.637 \\
    \hline
    \end{tabularx}
    \caption{Classification performance for \textbf{DDH} grading on X-ray images}
    \label{tab:classification_results}
\end{table}

\textbf{Note:} For the Grade 1 vs. Grades 2-4 classification analysis, cases where Retuve returned a result of "0" were logically classified as IHDI Grade 2 or higher, as a "0" result represents a Retuve processing error and indicates the system's inability to confidently classify the case as normal (Grade 1).

\begin{figure}[H]
    \centering
    \includegraphics[width=\textwidth]{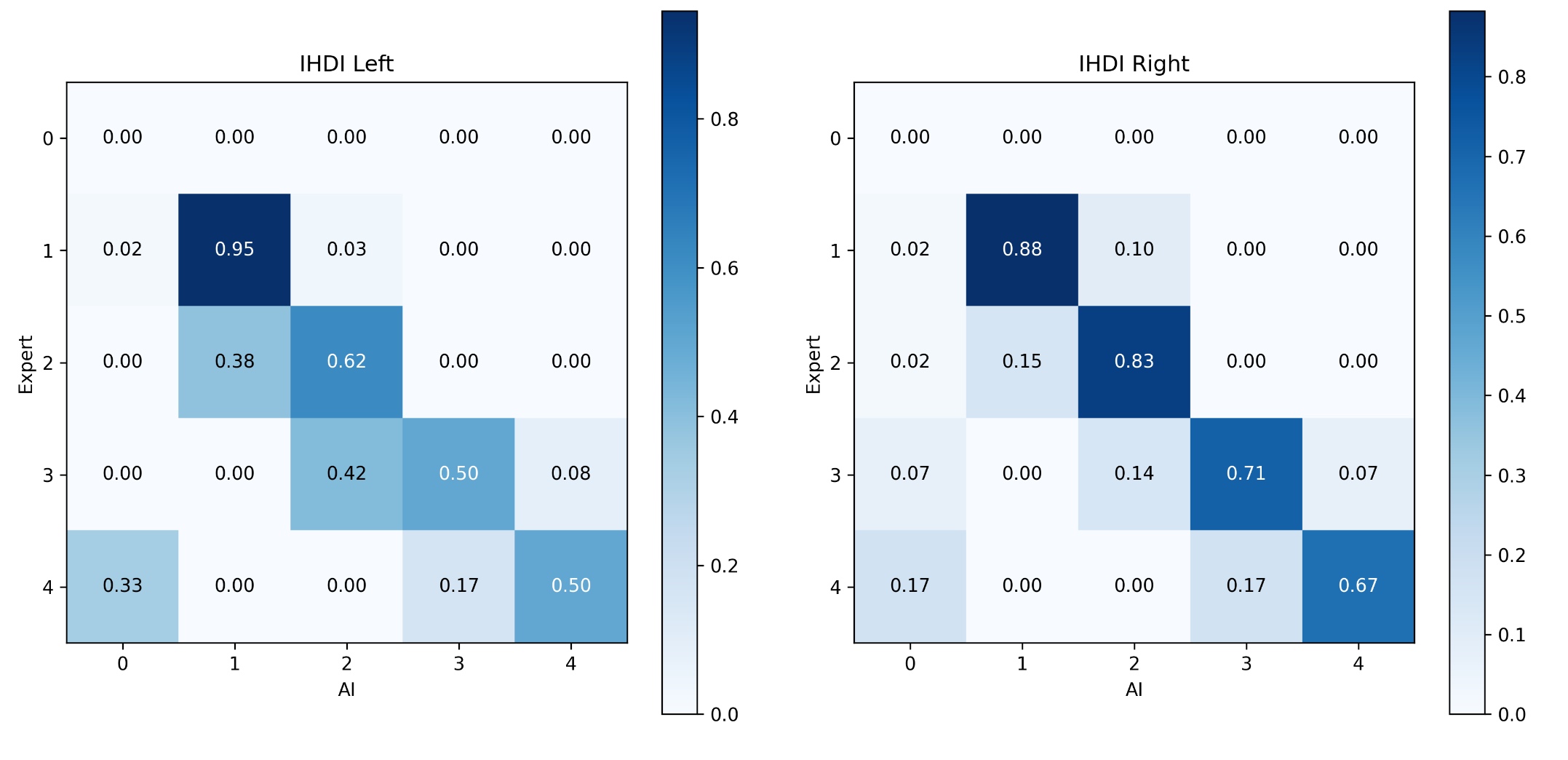}
    \caption{X-ray IHDI classification confusion matrices. Left: Left hip confusion matrix. Right: Right hip confusion matrix.}
    \label{fig:xray_confusion}
\end{figure}

\subsubsection{Discussion of Results}

Ultrasound alpha angle measurements show contrasting ICC values: low absolute agreement (0.47) but high consistency (0.86), suggesting systematic offset requiring threshold calibration rather than traditional Graf thresholds. Coverage measurements demonstrate strong performance in both absolute agreement (0.88) and consistency (0.92). This is just below agreements shown in previous work \cite{HAREENDRANATHAN201689}. Detailed scatter plots showing the correlation between expert and AI measurements are provided in \textbf{Figure \ref{fig:ultrasound_validation}}.

X-ray analysis demonstrates excellent reliability with ICC values ranging from 0.845-0.902 for both acetabular and Wilberg indices. Classification performance for Grade 1 vs. higher-grade IHDI (F1: 0.940) demonstrates screening utility. These metrics are within the range found in previous work \cite{icc_acetabular, icc_wilberg}. Correlation scatter plots for both acetabular index and Wilberg angle measurements are shown in \textbf{Figure \ref{fig:acetabular_scatter}} and \textbf{Figure \ref{fig:wilberg_scatter}}, respectively.

We show a comparison of Retuve's ICC performance with published literature values in \textbf{Table \ref{tab:literature_comparison}}.

\subsubsection{Performance Comparison with Literature}

\begin{table}[H]
    \centering
    \small
    \begin{tabularx}{\textwidth}{|>{\raggedright\arraybackslash}p{3.5cm}|>{\centering\arraybackslash}X|>{\centering\arraybackslash}X|>{\centering\arraybackslash}X|}
    \hline
    \textbf{Metric} & \textbf{Retuve ICC} & \textbf{Literature ICC} & \textbf{Performance} \\
    \hline
    \multicolumn{4}{|l|}{\textbf{Ultrasound}} \\
    \hline
    Alpha Angle (Consistency) & 0.86 & 0.87-0.92 \cite{HAREENDRANATHAN201689} & Comparable \\
    \hline
    \multicolumn{4}{|l|}{\textbf{X-ray}} \\
    \hline
    Acetabular Index & 0.845-0.860 & 0.811-0.996 \cite{icc_acetabular} & Within range \\
    \hline
    Wilberg Index & 0.891-0.902 & 0.918 \cite{icc_wilberg} & Comparable \\
    \hline
    \end{tabularx}
    \caption{Comparison of Retuve's ICC performance with published literature values. \textbf{Important Limitation:} Direct comparisons are challenging due to different datasets, imaging protocols, expert populations, and measurement methodologies across studies. Coverage was excluded as no literature values were found.}
    \label{tab:literature_comparison}
\end{table}

\textbf{Limitations:} The ultrasound dataset is small (172 images) and both datasets are single-center, potentially limiting generalizability. As the first fully reproducible \textbf{DDH} framework, direct comparisons with other models are challenging due to lack of available code/data in previous studies. Multi-center validation and comprehensive benchmarking remain important future work.

\subsubsection{Error Analysis}

Ultrasound validation identified outliers: 5 over-marked alpha angle cases and 3 coverage cases, attributed to poor scan quality or expert labeling errors. Results demonstrate Retuve's multi-modality \textbf{DDH} analysis capability with strong ultrasound coverage consistency and robust X-ray classification performance.

\section{Limitations and Future Development}

\subsection{Development Roadmap}

We show a development roadmap with links to the relevant GitHub issues where applicable.

\begin{table}[H]
    \centering
    \small
    \begin{tabularx}{\textwidth}{|>{\raggedright\arraybackslash}p{2cm}|>{\raggedright\arraybackslash}p{3cm}|>{\raggedright\arraybackslash}X|>{\centering\arraybackslash}p{1.5cm}|}
    \hline
    \textbf{Category} & \textbf{Feature/Goal} & \textbf{Description} & \textbf{Timeline} \\
    \hline
    \multirow{2}{3cm}{Platform\\Integration} & \href{https://github.com/radoss-org/retuve/issues/106}{PACS-AI Integration} & Direct PACS integration for automated \textbf{DDH} workflow & Q4 2025 - Q1 2026 \\
    \cline{2-4}
    & \href{https://github.com/radoss-org/retuve/issues/8}{ChRIS Plugin} & Research platform integration & Q3 2025 \\
    \hline
    \multirow{2}{3cm}{AI\\Enhancement} & \href{https://github.com/radoss-org/retuve/issues/56}{Scan Quality AI} & Real-time quality assessment for point-of-care & Q3 2025 \\
    \cline{2-4}
    & \href{https://github.com/radoss-org/retuve/issues/49}{Advanced Detection} & Comprehensive \textbf{DDH} classification beyond angles & Q4 2025 \\
    \hline
    \multirow{3}{3cm}{Clinical\\Research} & Retrospective Studies & Large-scale \textbf{DDH} detection validation & Q2-Q4 2026 \\
    \cline{2-4}
    & Retrospective Natural History Analysis & Treatment necessity prediction models & Q4 2025 - Q1 2026 \\
    \cline{2-4}
    & Multi-Center Studies & Multi-institutional validation & Q3-Q4 2026 \\
    \hline
    \multirow{2}{3cm}{Framework\\Enhancement} & \href{https://github.com/radoss-org/retuve/issues/32}{Custom Metrics} & User-defined parameters & Q1-Q2 2026 \\
    \cline{2-4}
    & \href{https://github.com/radoss-org/open-hip-dysplasia/issues/2}{Dataset Expansion} & Additional annotated datasets & Ongoing \\
    \hline
    \multirow{2}{3cm}{Clinical\\Validation} & Clinical Study & Prospective validation study & TBD \\
    \cline{2-4}
    & Usability Study & Clinical workflow assessment & TBD \\
    \hline
    \end{tabularx}
    \caption{Retuve Development Roadmap for 2025-2026}
    \label{tab:roadmap}
\end{table}

\subsection{Limitations}

This study has several important limitations. The validation datasets are relatively small, with only 172 ultrasound images and 1,878 X-ray images after processing, limiting generalizability and contributing to wider confidence intervals for some measurements. Both datasets represent single-center collections, which may introduce institutional bias and limit broader applicability across different imaging protocols, equipment vendors, and patient populations. The ultrasound alpha angle measurements show systematic offset requiring calibration, indicating that current Graf method thresholds may not be directly applicable without adjustment. No clinical validation or usability studies have been conducted to assess real-world diagnostic accuracy, clinical impact, or user experience in actual healthcare settings. As the first fully reproducible open-source DDH framework, direct performance comparisons with other AI models are challenging due to lack of available code and datasets in previous studies. The current release focuses on single-frame analysis and does not yet incorporate dynamic ultrasound sweep analysis or advanced quality assessment features that may be critical for robust clinical deployment. Multi-center validation, clinical trials, and comprehensive usability studies remain essential future work to establish clinical utility and safety.

\section{Conclusion}
We introduced Retuve, the first fully open-source \textbf{DDH} collaborative framework supporting X-ray and ultrasound imaging with open datasets, models, and Python API for reproducible development. The modular architecture enables easy AI model integration and features novel landmark-based X-ray analysis.

Future development will enhance robustness for challenging scenarios including \textbf{2DUS} sweeps and poor-quality images, expand expert-annotated datasets, and investigate landmark-based ultrasound approaches.

By democratizing \textbf{DDH} diagnosis through open-source AI, Retuve empowers global collaborative advancement. We encourage community participation in benchmarking, contributing models, and expanding capabilities to accelerate early \textbf{DDH} detection and improve infant outcomes while reducing long-term complications. Retuve represents a significant step toward accessible, transparent healthcare solutions and a model for medical imaging AI tools.

\section{Acknowledgements}

We would like to specifically thank Abhilash Hareendranathan for inspiring the paper format. We thank CIFAR, Alberta Innovates, The Arthritis Society, the TD Ready award, WCHRI, and CIHR for funding support. We would also like to thank \href{https://apps.ualberta.ca/directory/person/maedeh3}{Mariana Dehghan} for providing reviews of the paper.

\section{Conflicts of Interest}
Adam McArthur is the CEO/Founder of RadOSS, a non-profit organization with the mission of developing and distributing open-source radiology software. A portion of Dr. Jaremko's academic time is supported by Medical Imaging Consultants, Edmonton. All other authors declare no conflicts of interest.

\bibliographystyle{retuve-arch}
\bibliography{references}

\appendix

\section{Validation Scatter Plots}

\subsection{Ultrasound Validation Results}

\begin{figure}[H]
    \centering
    \includegraphics[width=\textwidth]{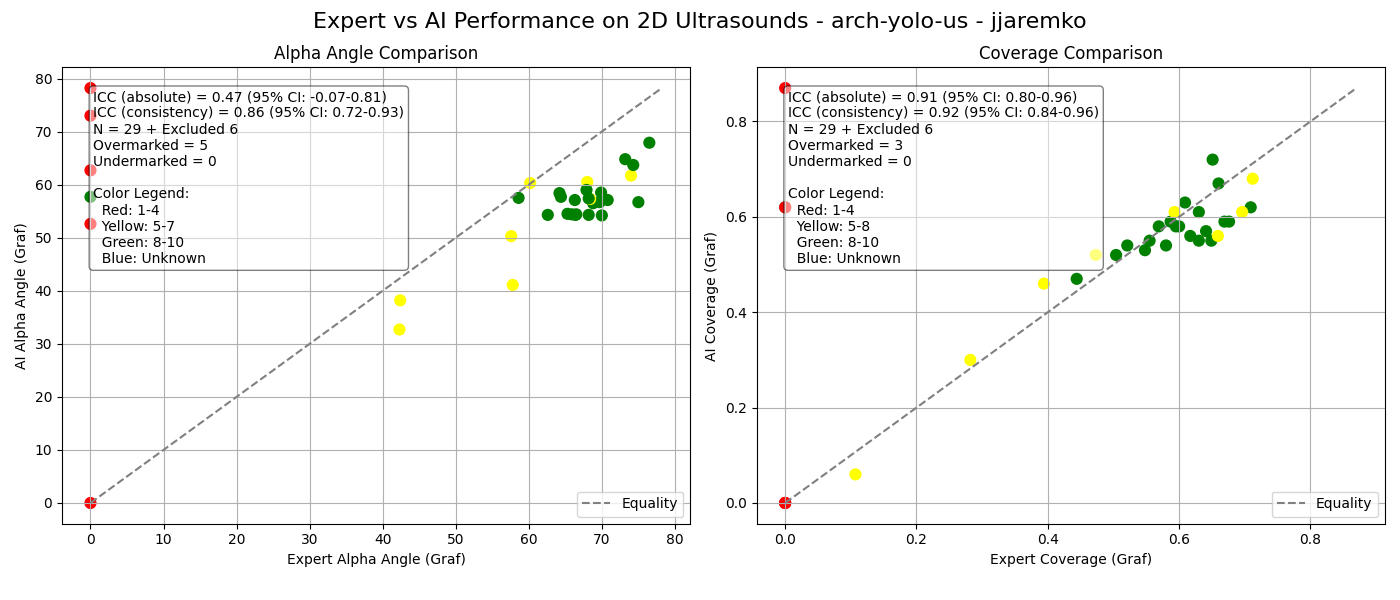}
    \caption{Ultrasound validation scatter plots. Left: Alpha angle correlation between expert and AI measurements. Right: Coverage correlation between expert and AI measurements.}
    \label{fig:ultrasound_validation}
\end{figure}

\subsection{X-ray Validation Results}

\begin{figure}[H]
    \centering
    \includegraphics[width=\textwidth]{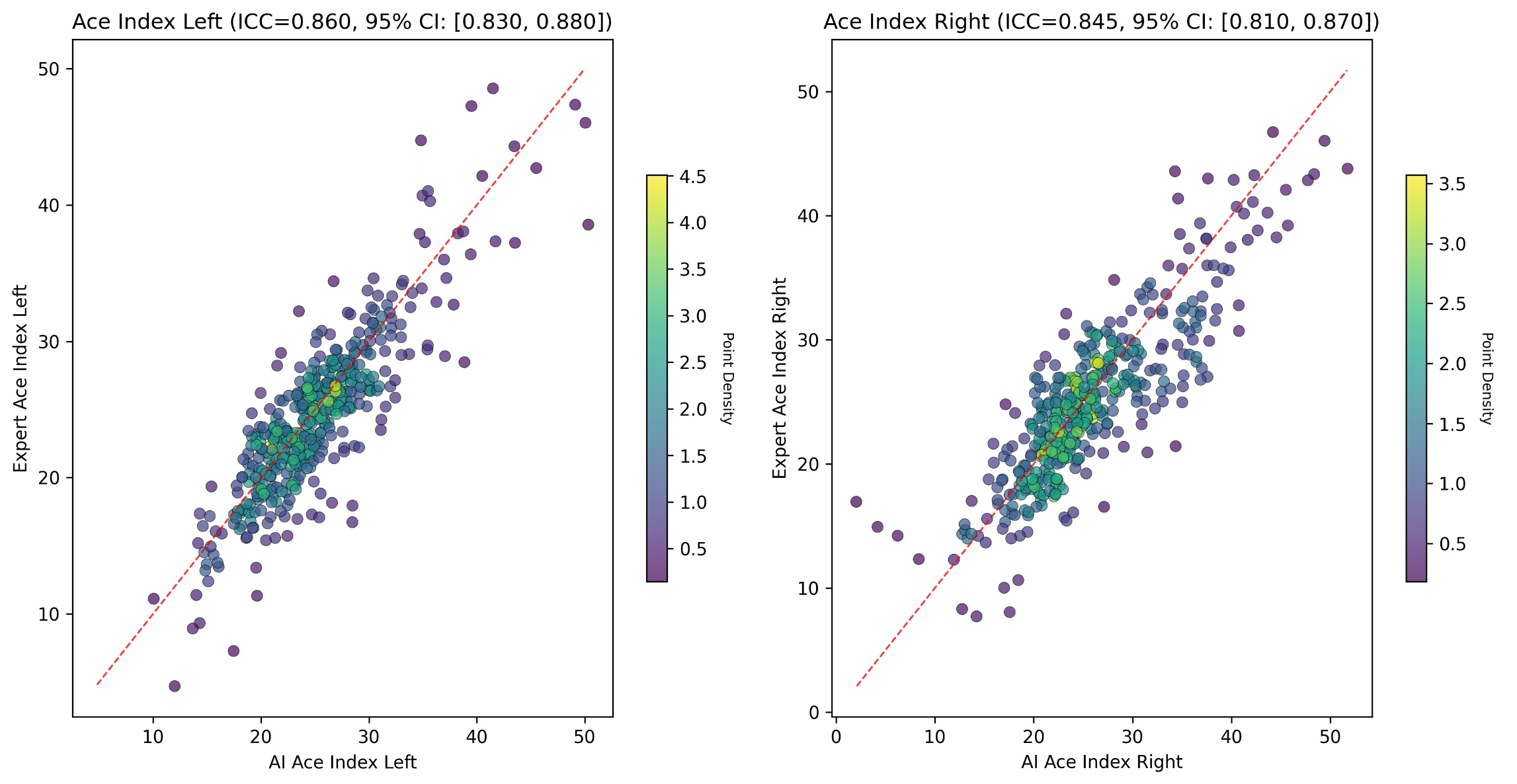}
    \caption{Acetabular index correlation scatter plots for left and right hips. Density is shown with a colour gradient from purple to yellow.}
    \label{fig:acetabular_scatter}
\end{figure}

\begin{figure}[H]
    \centering
    \includegraphics[width=\textwidth]{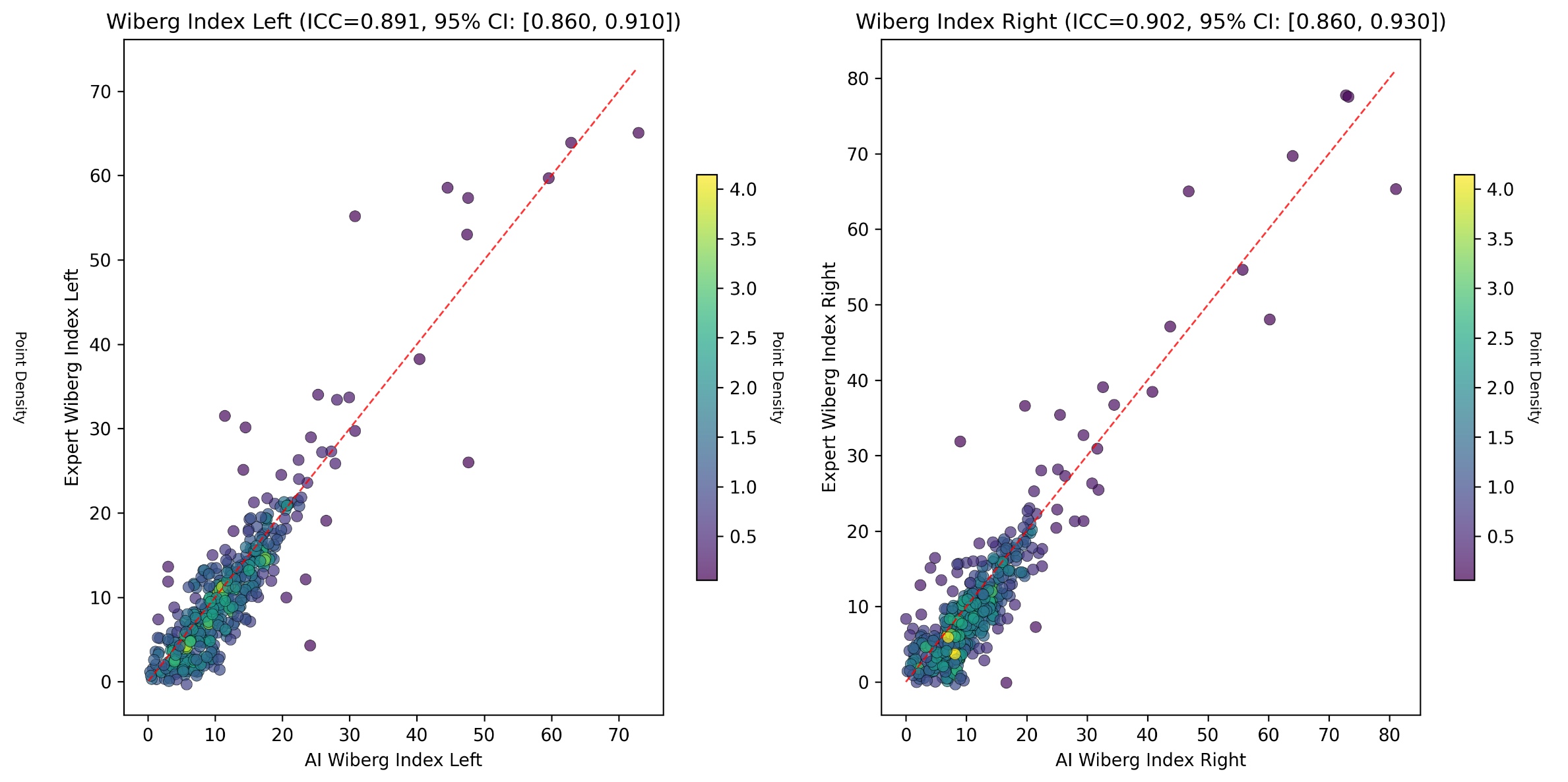}
    \caption{Wilberg angle correlation scatter plots for left and right hips. Density is shown with a colour gradient from purple to yellow.}
    \label{fig:wilberg_scatter}
\end{figure}

\end{document}